\documentclass[aps,prl,twocolumn,superscriptaddress,showpacs]{revtex4-1}

\usepackage{graphicx}

\begin{document}

\title{Unraveling orbital hybridization of triplet emitters at
the metal-organic interface}

\author{Pascal R. Ewen}
\affiliation{Physikalisches Institut and Center for Nanotechnology
(CeNTech), Westf\"alische Wilhelms-Universit\"at M\"unster, 48149
M\"unster, Germany}

\author{Jan Sanning}
\affiliation{Physikalisches Institut and Center for Nanotechnology
(CeNTech), Westf\"alische Wilhelms-Universit\"at M\"unster, 48149
M\"unster, Germany}

\author{Nikos L. Doltsinis}
\affiliation{Institut f\"ur Festk\"orpertheorie, Westf\"alische
Wilhelms-Universit\"at M\"unster, 48149 M\"unster, Germany}

\author{Matteo Mauro}
\affiliation{Laboratoire de Chimie et des Biomat\'eriaux
Supramol\'eculaires, Institut de Science et d'Ing\'enierie
Supramol\'eculaires (ISIS), Universit\'e de Strasbourg, 67083
Strasbourg, France}

\author{Cristian A. Strassert}
\email{ca.s@uni-muenster.de} \affiliation{Physikalisches Institut
and Center for Nanotechnology (CeNTech), Westf\"alische
Wilhelms-Universit\"at M\"unster, 48149 M\"unster, Germany}

\author{Daniel Wegner}
\email{daniel.wegner@uni-muenster.de} \affiliation{Physikalisches
Institut and Center for Nanotechnology (CeNTech), Westf\"alische
Wilhelms-Universit\"at M\"unster, 48149 M\"unster, Germany}

\date{November 14, 2013}

\begin{abstract}
We have investigated the structural and electronic properties of
phosphorescent planar platinum(II) complexes at the interface of
Au(111) with submolecular resolution using combined scanning
tunneling microscopy and spectroscopy as well as density functional
theory. Our analysis shows that molecule-substrate coupling and
lateral intermolecular interactions are weak. While the ligand
orbitals remain essentially unchanged upon contact to the substrate,
we found modified electronic behavior at the Pt atom due to local
hybridization and charge transfer to the substrate. Thus, this novel
class of phosphorescent molecules exhibits well-defined and tunable
interaction with its local environment.
\end{abstract}

\pacs{68.37.Ef, 68.55.-a, 73.20.Hb, 81.07.Pr}

\maketitle


Organic light emitting diodes (OLEDs) are currently investigated
extensively as alternative, highly efficient lighting sources and
for display technologies \cite{Muellen2006}. While pure organic
emitters are limited to a quantum efficiency of 25\% by fluorescence
from excited singlet states, the introduction of a heavy-metal atom
with large spin-orbit coupling can increase the efficiency up to
100\% by triplet harvesting and phosphorescence
\cite{Baldo1998,Yersin2007}. By far, most research and applications
have concentrated on iridium(III) complexes that require octahedral
coordination of the Ir atom \cite{Flamigni2007}. Interactions of an
Ir complex with its local environment lack defined directionality
and are thus barely controllable, which usually leads to quenching
effects and reduced quantum efficiencies when the Ir-complex loading
in an OLED is too high \cite{Reineke2009b}. In contrast, recently
synthesized platinum(II) complexes not only yield quantum
efficiencies of more than 85\%, but they also exhibit no quenching
effects even when aggregated into fibers or gels
\cite{Strassert2011,Mydlak2011}. The planar geometry enables
well-defined interactions with the local environment that should be
tunable.

The electronic properties of triplet emitters at conductive
interfaces is fundamental for the understanding of
electroluminescent devices, in particular light-emitting
electrochemical cells (LEECs) and OLEDs. Scanning tunneling
microscopy (STM) is an ideal tool to study single molecules in a
conductive environment with high spatial resolution, while
spectroscopic mapping via scanning tunneling spectroscopy (STS) can
identify energetic positions and spatial distributions of molecular
frontier orbitals
\cite{Lu2003,Repp2005,Wegner2008,Kahng2009,Barth2009}. Surprisingly,
so far only few studies have used scanning probe techniques to study
triplet emitters at the nanoscale
\cite{Oncel2008,Gersen2006,Ng2009}, and none of them have utilized
the advantages of combined STM and STS.

We report on a combined STM and STS study of two Pt-based triplet
emitters at the interface of Au(111). We found that the molecules
self-assemble into densely-packed monolayers. Through local STS
spectra and energy-resolved spectroscopic maps we identified various
occupied and unoccupied frontier orbitals. Comparison with density
functional theory (DFT) reveals that intermolecular interactions as
well as the coupling of the ligand to the substrate are relatively
weak. However, we found modified electronic behavior at the Pt atom,
which we attribute to hybridization with the Au surface leading to
partial depopulation of the Pt $d_{z^2}$-orbital. The molecular
coupling with its local environment is thus well-defined and offers
opportunities for controlled tuning of electronic and optical
properties of this type of phosphorescent complexes.


All experiments were performed under ultra-high vacuum (UHV)
conditions (base pressure $1\cdot10^{-10}\,$mbar) using a commercial
low-temperature STM (Createc LT-STM). The synthesis of pyridine
2,6-bis(3-(tri\-fluoro\-methyl)-1,2,4-tri\-azolato-5-yl)py\-ri\-dine
platinum(II) (Pt-H) and 4-pentyl-pyridine
2,6-bis(3-(tri\-fluoro\-methyl)-1,2,4-tri\-azolato-5-yl)py\-ri\-dine
platinum(II) (Pt-amyl) is described elsewhere
\cite{patent12,Mydlak2011}. The Au(111) single crystal was cleaned
by standard sputter-annealing procedures, followed by thermal
deposition of either Pt-H or Pt-amyl from a Knudsen cell at about
480~K and 420~K, respectively, while keeping the Au substrate at
room temperature. The sample was then transferred \textit{in situ}
into the cold STM (sample temperature 5~K). Topography images were
taken in constant-current mode. For STS spectra, the differential
conductance $dI/dV$ was measured as a function of the sample bias
$V$ using the lock-in technique under open feedback conditions
(typical modulation: 10--40~mV$_{\text{rms}}$). Energy-resolved
spectral maps were acquired by measuring $dI/dV$ at fixed bias while
scanning the surface in constant-current mode. For the DFT
calculations of Pt-H, Kohn-Sham molecular orbitals were calculated
in the gas phase with the Gaussian~09 package using the B3LYP
functional and the SDD basis set \cite{g09-short,b3lyp,andrae90}.

%
%

\begin{figure}
\begin{center}
\includegraphics[width=\columnwidth]{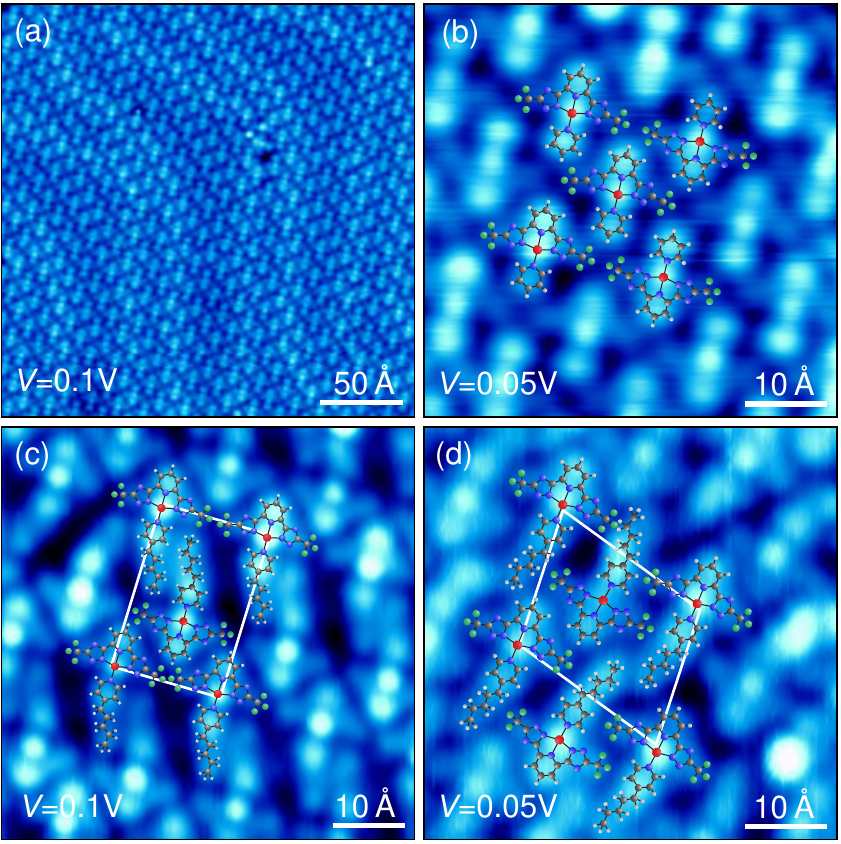}
\caption{\label{fig1} (Color Online). (a,b)~Overview and close-up
STM image of the monolayer of Pt-H, respectively. (c,d)~Close-up
images of two different packing structures of Pt-amyl. While the
presence of additional amyl groups in Pt-amyl lead to formation of
pairs and an overall high degree of order, Pt-H exhibits randomly
distributed parallel or antiparallel orientation (d). The overlaid
molecular structure models demonstrate that the high spatial
resolution enables an unambiguous identification of the molecules.}
\end{center}
\end{figure}

We first focus on the structural characterization of the as-grown
molecular films. Figure~\ref{fig1} shows STM images of Pt-H (a,b)
and Pt-amyl (c,d) on Au(111), respectively. We found that all
molecules are intact and self-assemble in densely packed monolayer
islands. The close-up views clearly show that the complexes can be
imaged with submolecular resolution, permitting identification by a
simple overlay of the structural models. The complexes consist of a
platinum atom surrounded by a tridentate ligand (TL) and an
ancillary ligand (AL) in square-planar coordination. The brightest
feature is a round protrusion in the center of the complexes that we
attribute to the Pt site. The TL shows up as an "arrow head" with
two lobes to the left and right of the Pt site (stemming from the
triazole-CF$_3$ groups) plus one above the Pt site (pyridine group).
The two complexes only differ in the AL, being a pyridine in case of
Pt-H and an amyl-pyridine for Pt-amyl. This difference can be
distinguished in the STM images: the C$_5$H$_11$ group is seen as an
additional "tail" in (c,d), whereas the bare pyridine in Pt-H only
appears as a round protrusion (b).

From a comparison of the monolayer structures we can deduce the
driving forces of self-assembly. Pt-H molecules are close-packed in
a rhombic lattice with side length 11.5(3)~{\AA} and an angle of
86(4)$^\circ$ (b). However, the molecular orientation at each
lattice point can either be parallel or antiparallel with respect to
the neighboring molecules. We did not find any nearest or
next-nearest neighbor correlations, i.e., the orientation is purely
random. The molecular symmetry axis is 12.0(5)$^\circ$ rotated
relative to the Au$\{11\bar{2}\}$ direction. While only one
self-assembled pattern is found for Pt-H, we observed two different
packings of Pt-amyl: at lower local coverage we found a rectangular
unit cell containing two molecules (c); at a nominally full
monolayer, we observed a denser oblique unit cell (d). The molecular
symmetry axes relative to Au$\{11\bar{2}\}$ are 9(3)$^\circ$ in (c)
as well as 4(3)$^\circ$ and 15(3)$^\circ$ in (d). We attribute the
more complex self-assembly of Pt-amyl to additional van der Waals
interactions between the amyl groups \cite{Bai2000,deFeyter2007}. On
the other hand, the orientational variations in Pt-H lead us to
conclude that there is no significant lateral interaction between
the TLs of neighboring molecules and that the structural arrangement
is simply a close-packing taking into account steric effects.
Furthermore, the various molecular angles relative to the Au(111)
substrate indicate that the molecule-substrate interaction is
relatively weak.

\begin{figure}
\begin{center}
\includegraphics[width=\columnwidth]{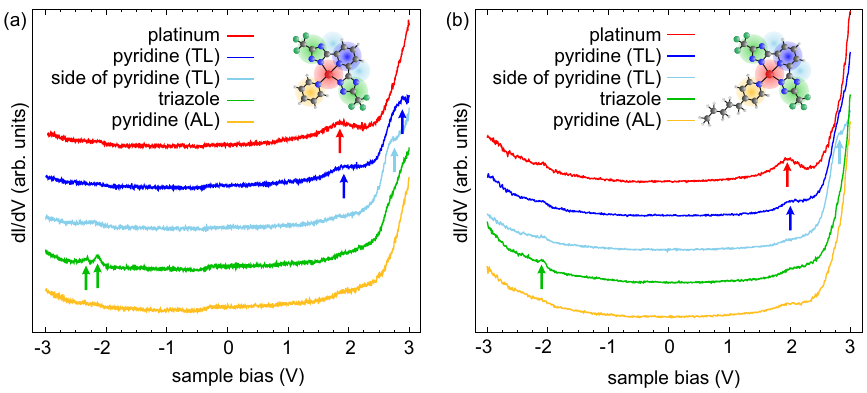}
\caption{\label{fig2} (Color Online). Local tunneling spectra of
Pt-H (a) and Pt-amyl (b) on Au(111). The spectra where taken over
the different parts of the molecule as indicated by the color code
of the molecular model. The arrows indicate features caused by
molecular orbitals.}
\end{center}
\end{figure}

Measurements of the differential conductance at different sites
within individual molecules permits to understand the local
electronic structure of the adsorbed Pt complexes, as $dI/dV$
variations reflect the spatial local density of states (LDOS)
distributions of molecular orbitals \cite{Lu2003}. Figure~\ref{fig2}
shows $dI/dV$ spectra for Pt-H and Pt-amyl, respectively. While
positions vary by some tens of mV for different molecules within the
monolayer, peak widths are in the order of 100~mV. Apart from a
systematic shift of about 0.1 eV, the spectra of Pt-H and Pt-amyl
are very similar. Occupied states can be found at about -2~eV,
mainly localized at the TL-triazole groups. Unoccupied states are
observed above about +2~eV at the TL-pyridine group.

\begin{figure}
\begin{center}
\includegraphics[width=\columnwidth]{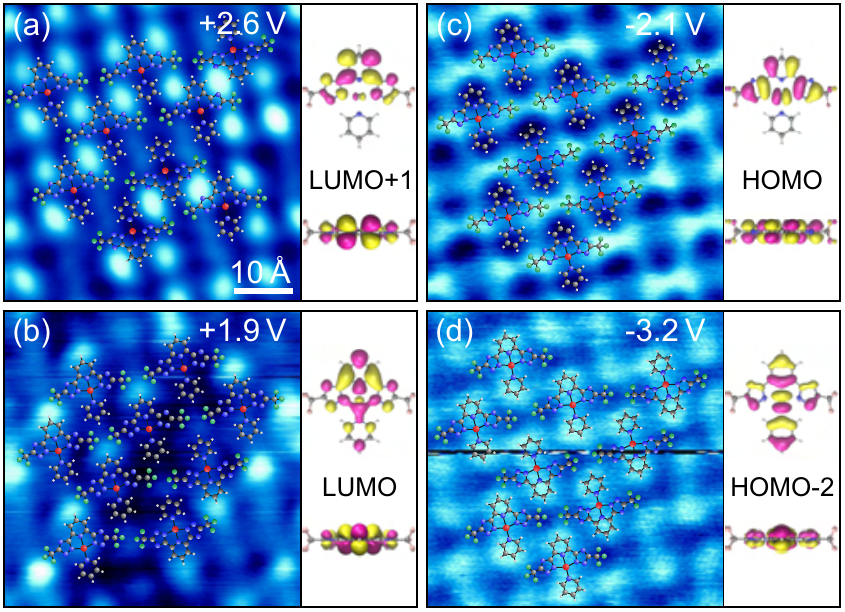}
\caption{\label{fig3} (Color Online). Series of $dI/dV$ maps
exhibiting LDOS at the ligands and assignment of corresponding
calculated orbitals of the gas-phase molecule. The unoccupied states
mainly show LDOS contributions from the TL-pyridine (a,b), while the
HOMO is localized at the triazole groups (c). The HOMO--2 exhibits
LDOS at both TL- and AL-pyridines. Note that the maps shown in (a,b)
have been acquired at a different position compared to those in
(c,d).}
\end{center}
\end{figure}

In order to connect the observed spatial LDOS variations to specific
molecular orbitals, we performed energy-resolved spectroscopic
mapping. As we found almost identical spectroscopic results for both
molecules, we restrict our discussion to Pt-H. We measured $dI/dV$
maps over a wide range from -3.2~V to +3.2~V. Figure~\ref{fig3}
summarizes maps at energies where we found contributions from the
ligand (i.e.\ not the Pt site). In order to correlate the observed
features with specific sites of the complex, the maps are
superimposed with molecular structures. We first describe the
unoccupied region. At voltages from 3.1~V down to 2.3~V, maps show
strong intensities at the left- and right-hand sides of the
TL-pyridine group of each molecule (Fig.~\ref{fig3}a). This is in
accordance with the spatial localization of the shoulder at 2.6~V in
STS spectra. Within the molecular layer, the highest intensities are
found in regions between two neighboring TL-pyridines and
significantly low intensity between two ALs. This proves that the
dominating signal stems from TL rather than the AL. The lack of
intensity at the center of the TL-pyridine suggests that the
underlying orbital exhibits a node in the mirror plane of this
group. At bias voltages below 2.3~V and ranging down to 1.6~V (b),
we observe a feature above the center of the TL-pyridine groups,
while no signal is found at the TL-triazoles or the ALs. Thus,
unoccupied ligand states are mainly confined to the TL-pyridine
group. In the occupied region, $dI/dV$ intensity is found at the
TL-triazole groups over a wide energy range from -1.8~V to -2.9~V,
whereas the pyridine groups of both TL and AL have almost no
intensity (c). This behavior is in good agreement with the spatial
inhomogeneity of peaks below -2.0~V in the tunneling spectra. Below
-3.0~V, the $dI/dV$ maps appear to be inverted as intensity is now
located at both AL- and TL-pyridine groups but not at the triazoles
(d).

The energy-resolved spectral maps can be assigned to particular
molecular orbitals via comparison with calculated orbitals of the
complex in the gas phase, as seen in Fig.~\ref{fig3}. The lowest
unoccupied molecular orbitals (LUMO and LUMO+1) are mainly localized
at the TL-pyridine group. While the LUMO is symmetric with respect
to the molecular mirror plane (b), the LUMO+1 is antisymmetric (a).
As the same symmetry is also found in the $dI/dV$ maps at about
2.0~V and 2.6~V, respectively, we assign the observed features to
the LUMO and LUMO+1, as shown in Fig.~\ref{fig3} \footnote{The
$dI/dV$ signal appears to be located next to rather than at the
molecule. This is merely an artifact when scanning in
constant-current mode.}. The calculated energy separation of 0.65~eV
for the gas-phase molecule is in good agreement with the
experimentally observed 0.7~eV. The calculated highest occupied
molecular orbital (HOMO) is mainly distributed across the triazole
groups of the TL. DFT predicts the HOMO-LUMO gap to be 3.88~eV,
which again agrees well with the peak separation of about 4~eV in
the tunneling spectra (Fig.~\ref{fig2}). The occupied spectroscopic
map in Fig.~\ref{fig3}d shows very good correspondence with the
calculated HOMO--2 that lies 0.68~eV lower than the HOMO. Overall,
the comparison shows that we can assign molecular orbitals to each
of the $dI/dV$ maps. We found that occupation, order and energy
separation of the calculated ligand orbitals in the gas phase are in
good agreement with the measured orbitals of the adsorbed complex,
i.e., the molecular properties at the ligand remain essentially
unchanged upon adsorption to the Au surface, again indicating an
overall weak molecule-substrate interaction.

\begin{figure}
\begin{center}
\includegraphics[width=\columnwidth]{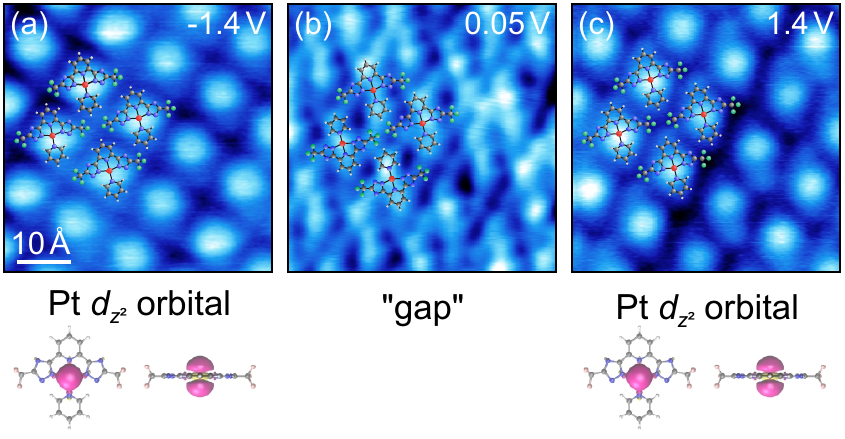}
\caption{\label{fig4} (Color Online). $dI/dV$ maps within the energy
windows -1.4 to 0.4~eV (a) and 1.0 to 1.8~eV (c) show localized LDOS
at the Pt atom. In between, the maps resemble the topography (b),
typical for a gap region. DFT predicts an occupied Pt
$d_{z^2}$-state for the gas-phase molecule. Its experimental
observation both below and above $E_F$ is indicative of a partially
depopulated orbital.}
\end{center}
\end{figure}

The situation is different when we look at the Pt site of the
complex. The DFT calculations show that the HOMO--1 is a $d_{z^2}$
orbital highly localized to the Pt atom (Fig.~\ref{fig4}). However,
no increased spectroscopic intensity is found at the Pt atom in any
of our $dI/dV$ maps below -2.0~V. Instead, we observe a
spectroscopic feature with increased LDOS intensity at the Pt atom
and no contributions from the ligand in maps between -1.4 and -0.4~V
(Fig.~\ref{fig4}a). Between -0.3 and 1.0~V, $dI/dV$ maps resemble
the topography (b), which is a typical observation for tunneling in
the region of the HOMO-LUMO gap of an adsorbed molecule (i.e.\ no
resonant tunneling into a distinct molecular orbital occurs).
Unexpectedly, we also found LDOS intensity localized at the Pt atom
in the unoccupied region between 1.0 and 1.8~V (c). At first glance,
this observation is surprising, because there can only be one
$d_{z^2}$ orbital at the Pt atom, and the calculations of the
gas-phase molecule predict it to be a fully occupied HOMO--1.

The modified electronic behavior of the complex at its Pt site can
be understood in terms of site-specific localized molecule-substrate
interaction and charge transfer. As can be seen in the side view of
the calculated molecular orbitals in Fig.~\ref{fig3} and \ref{fig4},
the Pt $d_{z^2}$-orbital extends out of the molecular plane farther
than the ligand orbitals. Therefore, locally increased overlap with
metallic states from the Au substrate can be expected at the Pt
site. Molecule-substrate hybridization can lead to a shift in
binding energy as well as a broadening of the $d_{z^2}$ state.
Strong overlap with the metallic substrate can also induce charge
transfer. We observed two orbitals at the Pt atom with identical
spatial distribution within the energy windows $-0.9 \pm 0.5$~eV and
$1.5 \pm 0.5$~eV, respectively. This behavior has been observed for
the case of a singly occupied molecular orbital (SOMO)
\cite{Repp2006}: at positive sample bias, a second electron can
tunnel into the SOMO as soon as the energy is large enough to
overcome the on-site Coulomb repulsion. This way, the SOMO is
detected both below and above the Fermi energy $E_F$. Typical
Coulomb energies for $d_{z^2}$ states in organometallic molecules
are around 2~eV \cite{Hou2005,Steinrueck2007,Mugarza2012}, in
accordance with the $2.4 \pm 0.7$~eV observed here. Another scenario
for our observation can be the formation of a bonding and an
antibonding state due to strong hybridization of the Pt
$d_{z^2}$-orbital with Au states \cite{Hoffmann88}. This latter
possibility can also explain why we do not observe these states as
peaks in our tunneling spectra.

At this point, we cannot distinguish between those two
possibilities. However, in both cases, the transfer of an electron
from the molecule to the substrate is required. Thus, we conclude
that the $d_{z^2}$ orbital of the Pt complex is only partially
occupied when adsorbed on the Au(111) surface. In contrast, the
ligand orbitals of the adsorbed molecule are unaffected by the
adsorption onto the Au surface. This means that charge transfer of
an electron between the molecule and the substrate only occurs
through the Pt atom, while the molecule-substrate interaction via
the ligand is small. We emphasize that our analysis of electronic
properties is identical for all complexes within the monolayer,
i.e., the molecules interact in a well-defined way with the
substrate. This is only possible due to its planar structure that
leads to distinct orbital overlaps. Hence, we expect that
well-defined interaction also occurs in host-guest environments as
well as within aggregated structures of Pt(II) complexes. This
interaction should even be tunable in a controlled fashion by
systematic variation of the ligand side groups (e.g., replacing
CF$_3$ by tert-butyl or adamantyl \cite{Mydlak2011}) or by changing
the host or substrate material \cite{Wegner2008,Harutyunyan2013}.

In summary, we have shown that Pt(II) complexes can be thermally
sublimated without any visible fragmentation, and they self-assemble
into well-ordered monolayers. STM and STS experiments were able to
probe the real-space and electronic structure with submolecular
resolution. The orbital structure of the ligands remains essentially
unaltered by the presence of the Au(111) surface. However, the
strongly changed electronic structure at the Pt atom provides
evidence for a localized charge transfer. These findings show that
-- contrary to the commonly used Ir(III) complexes -- planar Pt(II)
complexes exhibit a well-defined directional interaction with their
local environment that offers the opportunity of engineering their
electronic and thereby also optical properties.

%
%
\begin{acknowledgments}
This work was supported by the Deutsche Forschungsgemeinschaft (DFG)
through project WE-4104/2-1 and the Transregional Collaborative
Research Center TRR 61, project B13.
\end{acknowledgments}

\bibliography{Platin_07}

\begin{thebibliography}{28}%
\makeatletter
\providecommand \@ifxundefined [1]{%
 \@ifx{#1\undefined}
}%
\providecommand \@ifnum [1]{%
 \ifnum #1\expandafter \@firstoftwo
 \else \expandafter \@secondoftwo
 \fi
}%
\providecommand \@ifx [1]{%
 \ifx #1\expandafter \@firstoftwo
 \else \expandafter \@secondoftwo
 \fi
}%
\providecommand \natexlab [1]{#1}%
\providecommand \enquote  [1]{``#1''}%
\providecommand \bibnamefont  [1]{#1}%
\providecommand \bibfnamefont [1]{#1}%
\providecommand \citenamefont [1]{#1}%
\providecommand \href@noop [0]{\@secondoftwo}%
\providecommand \href [0]{\begingroup \@sanitize@url \@href}%
\providecommand \@href[1]{\@@startlink{#1}\@@href}%
\providecommand \@@href[1]{\endgroup#1\@@endlink}%
\providecommand \@sanitize@url [0]{\catcode `\\12\catcode `\$12\catcode
  `\&12\catcode `\#12\catcode `\^12\catcode `\_12\catcode `\%12\relax}%
\providecommand \@@startlink[1]{}%
\providecommand \@@endlink[0]{}%
\providecommand \url  [0]{\begingroup\@sanitize@url \@url }%
\providecommand \@url [1]{\endgroup\@href {#1}{\urlprefix }}%
\providecommand \urlprefix  [0]{URL }%
\providecommand \Eprint [0]{\href }%
\providecommand \doibase [0]{http://dx.doi.org/}%
\providecommand \selectlanguage [0]{\@gobble}%
\providecommand \bibinfo  [0]{\@secondoftwo}%
\providecommand \bibfield  [0]{\@secondoftwo}%
\providecommand \translation [1]{[#1]}%
\providecommand \BibitemOpen [0]{}%
\providecommand \bibitemStop [0]{}%
\providecommand \bibitemNoStop [0]{.\EOS\space}%
\providecommand \EOS [0]{\spacefactor3000\relax}%
\providecommand \BibitemShut  [1]{\csname bibitem#1\endcsname}%
\let\auto@bib@innerbib\@empty
\bibitem [{\citenamefont {M{\"u}llen}\ and\ \citenamefont
  {Scherf}(2006)}]{Muellen2006}%
  \BibitemOpen
  \bibinfo {editor} {\bibfnamefont {K.}~\bibnamefont {M{\"u}llen}}\ and\
  \bibinfo {editor} {\bibfnamefont {U.}~\bibnamefont {Scherf}},\ eds.,\
  \href@noop {} {\emph {\bibinfo {title} {Organic Light-Emitting Devices}}}\
  (\bibinfo  {publisher} {Wiley-VCH},\ \bibinfo {address} {Weinheim},\ \bibinfo
  {year} {2006})\BibitemShut {NoStop}%
\bibitem [{\citenamefont {Baldo}\ \emph {et~al.}(1998)\citenamefont {Baldo},
  \citenamefont {O'Brien}, \citenamefont {Shoustikov}, \citenamefont {Sibley},
  \citenamefont {Thomson},\ and\ \citenamefont {Forrest}}]{Baldo1998}%
  \BibitemOpen
  \bibfield  {author} {\bibinfo {author} {\bibfnamefont {M.~A.}\ \bibnamefont
  {Baldo}}, \bibinfo {author} {\bibfnamefont {D.}~\bibnamefont {O'Brien}},
  \bibinfo {author} {\bibfnamefont {A.}~\bibnamefont {Shoustikov}}, \bibinfo
  {author} {\bibfnamefont {S.}~\bibnamefont {Sibley}}, \bibinfo {author}
  {\bibfnamefont {M.~E.}\ \bibnamefont {Thomson}}, \ and\ \bibinfo {author}
  {\bibfnamefont {S.~R.}\ \bibnamefont {Forrest}},\ }\href@noop {} {\bibfield
  {journal} {\bibinfo  {journal} {Nature}\ }\textbf {\bibinfo {volume} {395}},\
  \bibinfo {pages} {151} (\bibinfo {year} {1998})}\BibitemShut {NoStop}%
\bibitem [{\citenamefont {Yersin}(2007)}]{Yersin2007}%
  \BibitemOpen
  \bibinfo {editor} {\bibfnamefont {H.}~\bibnamefont {Yersin}},\ ed.,\
  \href@noop {} {\emph {\bibinfo {title} {Highly Efficient OLEDs with
  Phosphorescent Materials}}}\ (\bibinfo  {publisher} {Wiley-VCH},\ \bibinfo
  {address} {Weinheim},\ \bibinfo {year} {2007})\BibitemShut {NoStop}%
\bibitem [{\citenamefont {Flamigni}\ \emph {et~al.}(2007)\citenamefont
  {Flamigni}, \citenamefont {Barbieri}, \citenamefont {Sabatini}, \citenamefont
  {Ventura},\ and\ \citenamefont {Barigelletti}}]{Flamigni2007}%
  \BibitemOpen
  \bibfield  {author} {\bibinfo {author} {\bibfnamefont {L.}~\bibnamefont
  {Flamigni}}, \bibinfo {author} {\bibfnamefont {A.}~\bibnamefont {Barbieri}},
  \bibinfo {author} {\bibfnamefont {C.}~\bibnamefont {Sabatini}}, \bibinfo
  {author} {\bibfnamefont {B.}~\bibnamefont {Ventura}}, \ and\ \bibinfo
  {author} {\bibfnamefont {F.}~\bibnamefont {Barigelletti}},\ }\href@noop {}
  {\bibfield  {journal} {\bibinfo  {journal} {Top. Curr. Chem.}\ }\textbf
  {\bibinfo {volume} {281}},\ \bibinfo {pages} {143} (\bibinfo {year}
  {2007})}\BibitemShut {NoStop}%
\bibitem [{\citenamefont {Reineke}\ \emph {et~al.}(2009)\citenamefont
  {Reineke}, \citenamefont {Schwartz}, \citenamefont {Walzer}, \citenamefont
  {Falke},\ and\ \citenamefont {Leo}}]{Reineke2009b}%
  \BibitemOpen
  \bibfield  {author} {\bibinfo {author} {\bibfnamefont {S.}~\bibnamefont
  {Reineke}}, \bibinfo {author} {\bibfnamefont {G.}~\bibnamefont {Schwartz}},
  \bibinfo {author} {\bibfnamefont {K.}~\bibnamefont {Walzer}}, \bibinfo
  {author} {\bibfnamefont {M.}~\bibnamefont {Falke}}, \ and\ \bibinfo {author}
  {\bibfnamefont {K.}~\bibnamefont {Leo}},\ }\href {\doibase
  http://dx.doi.org/10.1063/1.3123815} {\bibfield  {journal} {\bibinfo
  {journal} {Appl. Phys. Lett.}\ }\textbf {\bibinfo {volume} {94}},\ \bibinfo
  {pages} {163305} (\bibinfo {year} {2009})}\BibitemShut {NoStop}%
\bibitem [{\citenamefont {Strassert}\ \emph {et~al.}(2011)\citenamefont
  {Strassert}, \citenamefont {Chien}, \citenamefont {{Galvez Lopez}},
  \citenamefont {Kourkoulos}, \citenamefont {Hertel}, \citenamefont
  {Meerholz},\ and\ \citenamefont {{De Cola}}}]{Strassert2011}%
  \BibitemOpen
  \bibfield  {author} {\bibinfo {author} {\bibfnamefont {C.~A.}\ \bibnamefont
  {Strassert}}, \bibinfo {author} {\bibfnamefont {C.-H.}\ \bibnamefont
  {Chien}}, \bibinfo {author} {\bibfnamefont {M.~D.}\ \bibnamefont {{Galvez
  Lopez}}}, \bibinfo {author} {\bibfnamefont {D.}~\bibnamefont {Kourkoulos}},
  \bibinfo {author} {\bibfnamefont {D.}~\bibnamefont {Hertel}}, \bibinfo
  {author} {\bibfnamefont {K.}~\bibnamefont {Meerholz}}, \ and\ \bibinfo
  {author} {\bibfnamefont {L.}~\bibnamefont {{De Cola}}},\ }\href@noop {}
  {\bibfield  {journal} {\bibinfo  {journal} {Angew. Chem. Int. Ed.}\ }\textbf
  {\bibinfo {volume} {50}},\ \bibinfo {pages} {946} (\bibinfo {year}
  {2011})}\BibitemShut {NoStop}%
\bibitem [{\citenamefont {Mydlak}\ \emph {et~al.}(2011)\citenamefont {Mydlak},
  \citenamefont {Mauro}, \citenamefont {Polo}, \citenamefont {Felicetti},
  \citenamefont {Leonhardt}, \citenamefont {Diener}, \citenamefont {{De
  Cola}},\ and\ \citenamefont {Strassert}}]{Mydlak2011}%
  \BibitemOpen
  \bibfield  {author} {\bibinfo {author} {\bibfnamefont {M.}~\bibnamefont
  {Mydlak}}, \bibinfo {author} {\bibfnamefont {M.}~\bibnamefont {Mauro}},
  \bibinfo {author} {\bibfnamefont {F.}~\bibnamefont {Polo}}, \bibinfo {author}
  {\bibfnamefont {M.}~\bibnamefont {Felicetti}}, \bibinfo {author}
  {\bibfnamefont {J.}~\bibnamefont {Leonhardt}}, \bibinfo {author}
  {\bibfnamefont {G.}~\bibnamefont {Diener}}, \bibinfo {author} {\bibfnamefont
  {L.}~\bibnamefont {{De Cola}}}, \ and\ \bibinfo {author} {\bibfnamefont
  {C.~A.}\ \bibnamefont {Strassert}},\ }\href@noop {} {\bibfield  {journal}
  {\bibinfo  {journal} {Chem. Mater.}\ }\textbf {\bibinfo {volume} {23}},\
  \bibinfo {pages} {3659} (\bibinfo {year} {2011})}\BibitemShut {NoStop}%
\bibitem [{\citenamefont {Lu}\ \emph {et~al.}(2003)\citenamefont {Lu},
  \citenamefont {Grobis}, \citenamefont {Khoo}, \citenamefont {Louie},\ and\
  \citenamefont {Crommie}}]{Lu2003}%
  \BibitemOpen
  \bibfield  {author} {\bibinfo {author} {\bibfnamefont {X.}~\bibnamefont
  {Lu}}, \bibinfo {author} {\bibfnamefont {M.}~\bibnamefont {Grobis}}, \bibinfo
  {author} {\bibfnamefont {K.~H.}\ \bibnamefont {Khoo}}, \bibinfo {author}
  {\bibfnamefont {S.~G.}\ \bibnamefont {Louie}}, \ and\ \bibinfo {author}
  {\bibfnamefont {M.~F.}\ \bibnamefont {Crommie}},\ }\href {\doibase
  10.1103/PhysRevLett.90.096802} {\bibfield  {journal} {\bibinfo  {journal}
  {Phys. Rev. Lett.}\ }\textbf {\bibinfo {volume} {90}},\ \bibinfo {pages}
  {096802} (\bibinfo {year} {2003})}\BibitemShut {NoStop}%
\bibitem [{\citenamefont {Repp}\ \emph {et~al.}(2005)\citenamefont {Repp},
  \citenamefont {Meyer}, \citenamefont {Stojkovi\ifmmode~\acute{c}\else
  \'{c}\fi{}}, \citenamefont {Gourdon},\ and\ \citenamefont
  {Joachim}}]{Repp2005}%
  \BibitemOpen
  \bibfield  {author} {\bibinfo {author} {\bibfnamefont {J.}~\bibnamefont
  {Repp}}, \bibinfo {author} {\bibfnamefont {G.}~\bibnamefont {Meyer}},
  \bibinfo {author} {\bibfnamefont {S.~M.}\ \bibnamefont
  {Stojkovi\ifmmode~\acute{c}\else \'{c}\fi{}}}, \bibinfo {author}
  {\bibfnamefont {A.}~\bibnamefont {Gourdon}}, \ and\ \bibinfo {author}
  {\bibfnamefont {C.}~\bibnamefont {Joachim}},\ }\href {\doibase
  10.1103/PhysRevLett.94.026803} {\bibfield  {journal} {\bibinfo  {journal}
  {Phys. Rev. Lett.}\ }\textbf {\bibinfo {volume} {94}},\ \bibinfo {pages}
  {026803} (\bibinfo {year} {2005})}\BibitemShut {NoStop}%
\bibitem [{\citenamefont {Wegner}\ \emph {et~al.}(2008)\citenamefont {Wegner},
  \citenamefont {Yamachika}, \citenamefont {Wang}, \citenamefont {Brar},
  \citenamefont {Bartlett}, \citenamefont {Long},\ and\ \citenamefont
  {Crommie}}]{Wegner2008}%
  \BibitemOpen
  \bibfield  {author} {\bibinfo {author} {\bibfnamefont {D.}~\bibnamefont
  {Wegner}}, \bibinfo {author} {\bibfnamefont {R.}~\bibnamefont {Yamachika}},
  \bibinfo {author} {\bibfnamefont {Y.}~\bibnamefont {Wang}}, \bibinfo {author}
  {\bibfnamefont {V.~W.}\ \bibnamefont {Brar}}, \bibinfo {author}
  {\bibfnamefont {B.~M.}\ \bibnamefont {Bartlett}}, \bibinfo {author}
  {\bibfnamefont {J.~R.}\ \bibnamefont {Long}}, \ and\ \bibinfo {author}
  {\bibfnamefont {M.~F.}\ \bibnamefont {Crommie}},\ }\href@noop {} {\bibfield
  {journal} {\bibinfo  {journal} {Nano Lett.}\ }\textbf {\bibinfo {volume}
  {8}},\ \bibinfo {pages} {131} (\bibinfo {year} {2008})}\BibitemShut {NoStop}%
\bibitem [{\citenamefont {Kim}\ \emph {et~al.}(2009)\citenamefont {Kim},
  \citenamefont {Son}, \citenamefont {Jang}, \citenamefont {Yoon},
  \citenamefont {Han},\ and\ \citenamefont {Kahng}}]{Kahng2009}%
  \BibitemOpen
  \bibfield  {author} {\bibinfo {author} {\bibfnamefont {H.}~\bibnamefont
  {Kim}}, \bibinfo {author} {\bibfnamefont {W.-j.}\ \bibnamefont {Son}},
  \bibinfo {author} {\bibfnamefont {W.~J.}\ \bibnamefont {Jang}}, \bibinfo
  {author} {\bibfnamefont {J.~K.}\ \bibnamefont {Yoon}}, \bibinfo {author}
  {\bibfnamefont {S.}~\bibnamefont {Han}}, \ and\ \bibinfo {author}
  {\bibfnamefont {S.-J.}\ \bibnamefont {Kahng}},\ }\href {\doibase
  10.1103/PhysRevB.80.245402} {\bibfield  {journal} {\bibinfo  {journal} {Phys.
  Rev. B}\ }\textbf {\bibinfo {volume} {80}},\ \bibinfo {pages} {245402}
  (\bibinfo {year} {2009})}\BibitemShut {NoStop}%
\bibitem [{\citenamefont {Weber-Bargioni}\ \emph {et~al.}(2008)\citenamefont
  {Weber-Bargioni}, \citenamefont {Auw\"arter}, \citenamefont {Klappenberger},
  \citenamefont {Reichert}, \citenamefont {Lefran{\c{c}}ois}, \citenamefont
  {Strunskus}, \citenamefont {W\"oll}, \citenamefont {Schiffrin}, \citenamefont
  {Pennec},\ and\ \citenamefont {Barth}}]{Barth2009}%
  \BibitemOpen
  \bibfield  {author} {\bibinfo {author} {\bibfnamefont {A.}~\bibnamefont
  {Weber-Bargioni}}, \bibinfo {author} {\bibfnamefont {W.}~\bibnamefont
  {Auw\"arter}}, \bibinfo {author} {\bibfnamefont {F.}~\bibnamefont
  {Klappenberger}}, \bibinfo {author} {\bibfnamefont {J.}~\bibnamefont
  {Reichert}}, \bibinfo {author} {\bibfnamefont {S.}~\bibnamefont
  {Lefran{\c{c}}ois}}, \bibinfo {author} {\bibfnamefont {T.}~\bibnamefont
  {Strunskus}}, \bibinfo {author} {\bibfnamefont {C.}~\bibnamefont {W\"oll}},
  \bibinfo {author} {\bibfnamefont {A.}~\bibnamefont {Schiffrin}}, \bibinfo
  {author} {\bibfnamefont {Y.}~\bibnamefont {Pennec}}, \ and\ \bibinfo {author}
  {\bibfnamefont {J.~V.}\ \bibnamefont {Barth}},\ }\href {\doibase
  10.1002/cphc.200700600} {\bibfield  {journal} {\bibinfo  {journal}
  {ChemPhysChem}\ }\textbf {\bibinfo {volume} {9}},\ \bibinfo {pages} {89}
  (\bibinfo {year} {2008})}\BibitemShut {NoStop}%
\bibitem [{\citenamefont {Oncel}\ and\ \citenamefont
  {Bernasek}(2008)}]{Oncel2008}%
  \BibitemOpen
  \bibfield  {author} {\bibinfo {author} {\bibfnamefont {N.}~\bibnamefont
  {Oncel}}\ and\ \bibinfo {author} {\bibfnamefont {S.~L.}\ \bibnamefont
  {Bernasek}},\ }\href@noop {} {\bibfield  {journal} {\bibinfo  {journal}
  {Appl. Phys. Lett.}\ }\textbf {\bibinfo {volume} {92}},\ \bibinfo {pages}
  {133305} (\bibinfo {year} {2008})}\BibitemShut {NoStop}%
\bibitem [{\citenamefont {Gersen}\ \emph {et~al.}(2006)\citenamefont {Gersen},
  \citenamefont {Schaub}, \citenamefont {Xu}, \citenamefont {Stensgaard},
  \citenamefont {Laegsgaard}, \citenamefont {Linderoth}, \citenamefont
  {Besenbacher}, \citenamefont {Nazeeruddin},\ and\ \citenamefont
  {Graetzel}}]{Gersen2006}%
  \BibitemOpen
  \bibfield  {author} {\bibinfo {author} {\bibfnamefont {H.}~\bibnamefont
  {Gersen}}, \bibinfo {author} {\bibfnamefont {R.}~\bibnamefont {Schaub}},
  \bibinfo {author} {\bibfnamefont {W.}~\bibnamefont {Xu}}, \bibinfo {author}
  {\bibfnamefont {I.}~\bibnamefont {Stensgaard}}, \bibinfo {author}
  {\bibfnamefont {E.}~\bibnamefont {Laegsgaard}}, \bibinfo {author}
  {\bibfnamefont {T.~R.}\ \bibnamefont {Linderoth}}, \bibinfo {author}
  {\bibfnamefont {F.}~\bibnamefont {Besenbacher}}, \bibinfo {author}
  {\bibfnamefont {M.~K.}\ \bibnamefont {Nazeeruddin}}, \ and\ \bibinfo {author}
  {\bibfnamefont {M.}~\bibnamefont {Graetzel}},\ }\href@noop {} {\bibfield
  {journal} {\bibinfo  {journal} {Appl. Phys. Lett.}\ }\textbf {\bibinfo
  {volume} {89}},\ \bibinfo {pages} {264102} (\bibinfo {year}
  {2006})}\BibitemShut {NoStop}%
\bibitem [{\citenamefont {Ng}\ \emph {et~al.}(2009)\citenamefont {Ng},
  \citenamefont {Loh}, \citenamefont {Li}, \citenamefont {Ho}, \citenamefont
  {Bai},\ and\ \citenamefont {Yip}}]{Ng2009}%
  \BibitemOpen
  \bibfield  {author} {\bibinfo {author} {\bibfnamefont {Z.}~\bibnamefont
  {Ng}}, \bibinfo {author} {\bibfnamefont {K.~P.}\ \bibnamefont {Loh}},
  \bibinfo {author} {\bibfnamefont {L.}~\bibnamefont {Li}}, \bibinfo {author}
  {\bibfnamefont {P.}~\bibnamefont {Ho}}, \bibinfo {author} {\bibfnamefont
  {P.}~\bibnamefont {Bai}}, \ and\ \bibinfo {author} {\bibfnamefont {J.~H.~K.}\
  \bibnamefont {Yip}},\ }\href@noop {} {\bibfield  {journal} {\bibinfo
  {journal} {ACS Nano}\ }\textbf {\bibinfo {volume} {3}},\ \bibinfo {pages}
  {2103} (\bibinfo {year} {2009})}\BibitemShut {NoStop}%
\bibitem [{\citenamefont {{De Cola}}\ \emph {et~al.}(2012)\citenamefont {{De
  Cola}}, \citenamefont {Strassert}, \citenamefont {Mathias}, \citenamefont
  {Mauro}, \citenamefont {Felicetti}, \citenamefont {Diener},\ and\
  \citenamefont {Leonhardt}}]{patent12}%
  \BibitemOpen
  \bibfield  {author} {\bibinfo {author} {\bibfnamefont {L.}~\bibnamefont {{De
  Cola}}}, \bibinfo {author} {\bibfnamefont {C.~A.}\ \bibnamefont {Strassert}},
  \bibinfo {author} {\bibfnamefont {M.}~\bibnamefont {Mathias}}, \bibinfo
  {author} {\bibfnamefont {M.}~\bibnamefont {Mauro}}, \bibinfo {author}
  {\bibfnamefont {M.}~\bibnamefont {Felicetti}}, \bibinfo {author}
  {\bibfnamefont {G.}~\bibnamefont {Diener}}, \ and\ \bibinfo {author}
  {\bibfnamefont {J.}~\bibnamefont {Leonhardt}},\ }\href@noop {} {}\bibinfo
  {howpublished} {Patent DE102011001007} (\bibinfo {year} {2012})\BibitemShut
  {NoStop}%
\bibitem [{g09()}]{g09-short}%
  \BibitemOpen
  \href@noop {} {}\bibinfo {note} {Gaussian 09, Revision A.1, M. J. Frisch et
  al., Gaussian, Inc., Pittsburgh PA, 2009.}\BibitemShut {Stop}%
\bibitem [{\citenamefont {Becke}(1993)}]{b3lyp}%
  \BibitemOpen
  \bibfield  {author} {\bibinfo {author} {\bibfnamefont {A.~D.}\ \bibnamefont
  {Becke}},\ }\href@noop {} {\bibfield  {journal} {\bibinfo  {journal} {J.
  Chem. Phys.}\ }\textbf {\bibinfo {volume} {98}},\ \bibinfo {pages} {5648}
  (\bibinfo {year} {1993})}\BibitemShut {NoStop}%
\bibitem [{\citenamefont {Andrae}\ \emph {et~al.}(1990)\citenamefont {Andrae},
  \citenamefont {H\"au{\ss}ermann}, \citenamefont {Dolg}, \citenamefont
  {Stoll},\ and\ \citenamefont {Preuss}}]{andrae90}%
  \BibitemOpen
  \bibfield  {author} {\bibinfo {author} {\bibfnamefont {D.}~\bibnamefont
  {Andrae}}, \bibinfo {author} {\bibfnamefont {U.}~\bibnamefont
  {H\"au{\ss}ermann}}, \bibinfo {author} {\bibfnamefont {M.}~\bibnamefont
  {Dolg}}, \bibinfo {author} {\bibfnamefont {H.}~\bibnamefont {Stoll}}, \ and\
  \bibinfo {author} {\bibfnamefont {H.}~\bibnamefont {Preuss}},\ }\href@noop {}
  {\bibfield  {journal} {\bibinfo  {journal} {Theor. Chem. Acc.}\ }\textbf
  {\bibinfo {volume} {77}},\ \bibinfo {pages} {123} (\bibinfo {year}
  {1990})}\BibitemShut {NoStop}%
\bibitem [{\citenamefont {Qiu}\ \emph {et~al.}(2000)\citenamefont {Qiu},
  \citenamefont {Wang}, \citenamefont {Yin}, \citenamefont {Zeng},
  \citenamefont {Xu},\ and\ \citenamefont {Bai}}]{Bai2000}%
  \BibitemOpen
  \bibfield  {author} {\bibinfo {author} {\bibfnamefont {X.}~\bibnamefont
  {Qiu}}, \bibinfo {author} {\bibfnamefont {C.}~\bibnamefont {Wang}}, \bibinfo
  {author} {\bibfnamefont {S.}~\bibnamefont {Yin}}, \bibinfo {author}
  {\bibfnamefont {Q.}~\bibnamefont {Zeng}}, \bibinfo {author} {\bibfnamefont
  {B.}~\bibnamefont {Xu}}, \ and\ \bibinfo {author} {\bibfnamefont
  {C.}~\bibnamefont {Bai}},\ }\href {\doibase 10.1021/jp993501j} {\bibfield
  {journal} {\bibinfo  {journal} {J. Phys. Chem. B}\ }\textbf {\bibinfo
  {volume} {104}},\ \bibinfo {pages} {3570} (\bibinfo {year}
  {2000})}\BibitemShut {NoStop}%
\bibitem [{\citenamefont {Furukawa}\ \emph {et~al.}(2007)\citenamefont
  {Furukawa}, \citenamefont {Tahara}, \citenamefont {De~Schryver},
  \citenamefont {Van~der Auweraer}, \citenamefont {Tobe},\ and\ \citenamefont
  {De~Feyter}}]{deFeyter2007}%
  \BibitemOpen
  \bibfield  {author} {\bibinfo {author} {\bibfnamefont {S.}~\bibnamefont
  {Furukawa}}, \bibinfo {author} {\bibfnamefont {K.}~\bibnamefont {Tahara}},
  \bibinfo {author} {\bibfnamefont {F.~C.}\ \bibnamefont {De~Schryver}},
  \bibinfo {author} {\bibfnamefont {M.}~\bibnamefont {Van~der Auweraer}},
  \bibinfo {author} {\bibfnamefont {Y.}~\bibnamefont {Tobe}}, \ and\ \bibinfo
  {author} {\bibfnamefont {S.}~\bibnamefont {De~Feyter}},\ }\href {\doibase
  10.1002/anie.200604782} {\bibfield  {journal} {\bibinfo  {journal} {Angew.
  Chem. Int. Ed.}\ }\textbf {\bibinfo {volume} {46}},\ \bibinfo {pages} {2831}
  (\bibinfo {year} {2007})}\BibitemShut {NoStop}%
\bibitem [{Note1()}]{Note1}%
  \BibitemOpen
  \bibinfo {note} {The $dI/dV$ signal appears to be located next to rather than
  at the molecule. This is merely an artifact when scanning in constant-current
  mode.}\BibitemShut {Stop}%
\bibitem [{\citenamefont {Repp}\ \emph {et~al.}(2006)\citenamefont {Repp},
  \citenamefont {Meyer}, \citenamefont {Paavilainen}, \citenamefont {Olsson},\
  and\ \citenamefont {Persson}}]{Repp2006}%
  \BibitemOpen
  \bibfield  {author} {\bibinfo {author} {\bibfnamefont {J.}~\bibnamefont
  {Repp}}, \bibinfo {author} {\bibfnamefont {G.}~\bibnamefont {Meyer}},
  \bibinfo {author} {\bibfnamefont {S.}~\bibnamefont {Paavilainen}}, \bibinfo
  {author} {\bibfnamefont {F.~E.}\ \bibnamefont {Olsson}}, \ and\ \bibinfo
  {author} {\bibfnamefont {M.}~\bibnamefont {Persson}},\ }\href {\doibase
  10.1126/science.1126073} {\bibfield  {journal} {\bibinfo  {journal}
  {Science}\ }\textbf {\bibinfo {volume} {312}},\ \bibinfo {pages} {1196}
  (\bibinfo {year} {2006})}\BibitemShut {NoStop}%
\bibitem [{\citenamefont {Zhao}\ \emph {et~al.}(2005)\citenamefont {Zhao},
  \citenamefont {Li}, \citenamefont {Chen}, \citenamefont {Xiang},
  \citenamefont {Wang}, \citenamefont {Pan}, \citenamefont {Wang},
  \citenamefont {Xiao}, \citenamefont {Yang}, \citenamefont {Hou},\ and\
  \citenamefont {Zhu}}]{Hou2005}%
  \BibitemOpen
  \bibfield  {author} {\bibinfo {author} {\bibfnamefont {A.}~\bibnamefont
  {Zhao}}, \bibinfo {author} {\bibfnamefont {Q.}~\bibnamefont {Li}}, \bibinfo
  {author} {\bibfnamefont {L.}~\bibnamefont {Chen}}, \bibinfo {author}
  {\bibfnamefont {H.}~\bibnamefont {Xiang}}, \bibinfo {author} {\bibfnamefont
  {W.}~\bibnamefont {Wang}}, \bibinfo {author} {\bibfnamefont {S.}~\bibnamefont
  {Pan}}, \bibinfo {author} {\bibfnamefont {B.}~\bibnamefont {Wang}}, \bibinfo
  {author} {\bibfnamefont {X.}~\bibnamefont {Xiao}}, \bibinfo {author}
  {\bibfnamefont {J.}~\bibnamefont {Yang}}, \bibinfo {author} {\bibfnamefont
  {J.~G.}\ \bibnamefont {Hou}}, \ and\ \bibinfo {author} {\bibfnamefont
  {Q.}~\bibnamefont {Zhu}},\ }\href {\doibase 10.1126/science.1113449}
  {\bibfield  {journal} {\bibinfo  {journal} {Science}\ }\textbf {\bibinfo
  {volume} {309}},\ \bibinfo {pages} {1542} (\bibinfo {year}
  {2005})}\BibitemShut {NoStop}%
\bibitem [{\citenamefont {Lukasczyk}\ \emph {et~al.}(2007)\citenamefont
  {Lukasczyk}, \citenamefont {Flechtner}, \citenamefont {Merte}, \citenamefont
  {Jux}, \citenamefont {Maier}, \citenamefont {Gottfried},\ and\ \citenamefont
  {Steinr\"uck}}]{Steinrueck2007}%
  \BibitemOpen
  \bibfield  {author} {\bibinfo {author} {\bibfnamefont {T.}~\bibnamefont
  {Lukasczyk}}, \bibinfo {author} {\bibfnamefont {K.}~\bibnamefont
  {Flechtner}}, \bibinfo {author} {\bibfnamefont {L.~R.}\ \bibnamefont
  {Merte}}, \bibinfo {author} {\bibfnamefont {N.}~\bibnamefont {Jux}}, \bibinfo
  {author} {\bibfnamefont {F.}~\bibnamefont {Maier}}, \bibinfo {author}
  {\bibfnamefont {J.~M.}\ \bibnamefont {Gottfried}}, \ and\ \bibinfo {author}
  {\bibfnamefont {H.-P.}\ \bibnamefont {Steinr\"uck}},\ }\href {\doibase
  10.1021/jp0652345} {\bibfield  {journal} {\bibinfo  {journal} {J. Phys. Chem.
  C}\ }\textbf {\bibinfo {volume} {111}},\ \bibinfo {pages} {3090} (\bibinfo
  {year} {2007})}\BibitemShut {NoStop}%
\bibitem [{\citenamefont {Mugarza}\ \emph {et~al.}(2012)\citenamefont
  {Mugarza}, \citenamefont {Robles}, \citenamefont {Krull}, \citenamefont
  {Koryt\'{a}r}, \citenamefont {Lorente},\ and\ \citenamefont
  {Gambardella}}]{Mugarza2012}%
  \BibitemOpen
  \bibfield  {author} {\bibinfo {author} {\bibfnamefont {A.}~\bibnamefont
  {Mugarza}}, \bibinfo {author} {\bibfnamefont {R.}~\bibnamefont {Robles}},
  \bibinfo {author} {\bibfnamefont {C.}~\bibnamefont {Krull}}, \bibinfo
  {author} {\bibfnamefont {R.}~\bibnamefont {Koryt\'{a}r}}, \bibinfo {author}
  {\bibfnamefont {N.}~\bibnamefont {Lorente}}, \ and\ \bibinfo {author}
  {\bibfnamefont {P.}~\bibnamefont {Gambardella}},\ }\href@noop {} {\bibfield
  {journal} {\bibinfo  {journal} {Phys. Rev. B}\ }\textbf {\bibinfo {volume}
  {85}},\ \bibinfo {pages} {155437} (\bibinfo {year} {2012})}\BibitemShut
  {NoStop}%
\bibitem [{\citenamefont {Hoffmann}(1988)}]{Hoffmann88}%
  \BibitemOpen
  \bibfield  {author} {\bibinfo {author} {\bibfnamefont {R.}~\bibnamefont
  {Hoffmann}},\ }\href@noop {} {\bibfield  {journal} {\bibinfo  {journal} {Rev.
  Mod. Phys.}\ }\textbf {\bibinfo {volume} {60}},\ \bibinfo {pages} {601}
  (\bibinfo {year} {1988})}\BibitemShut {NoStop}%
\bibitem [{\citenamefont {Harutyunyan}\ \emph {et~al.}(2013)\citenamefont
  {Harutyunyan}, \citenamefont {Callsen}, \citenamefont {Allmers},
  \citenamefont {Caciuc}, \citenamefont {Bl{\"u}gel}, \citenamefont
  {Atodiresei},\ and\ \citenamefont {Wegner}}]{Harutyunyan2013}%
  \BibitemOpen
  \bibfield  {author} {\bibinfo {author} {\bibfnamefont {H.}~\bibnamefont
  {Harutyunyan}}, \bibinfo {author} {\bibfnamefont {M.}~\bibnamefont
  {Callsen}}, \bibinfo {author} {\bibfnamefont {T.}~\bibnamefont {Allmers}},
  \bibinfo {author} {\bibfnamefont {V.}~\bibnamefont {Caciuc}}, \bibinfo
  {author} {\bibfnamefont {S.}~\bibnamefont {Bl{\"u}gel}}, \bibinfo {author}
  {\bibfnamefont {N.}~\bibnamefont {Atodiresei}}, \ and\ \bibinfo {author}
  {\bibfnamefont {D.}~\bibnamefont {Wegner}},\ }\href {\doibase
  10.1039/C3CC42574F} {\bibfield  {journal} {\bibinfo  {journal} {Chem.
  Commun.}\ }\textbf {\bibinfo {volume} {49}},\ \bibinfo {pages} {5993}
  (\bibinfo {year} {2013})}\BibitemShut {NoStop}%
\end{thebibliography}%

\end{document}